\documentclass{article}
\usepackage[T1]{fontenc}
\usepackage{geometry}
\usepackage{graphics}

\makeatletter

\providecommand{\LyX}{L\kern-.1667em\lower.25em\hbox{Y}\kern-.125emX\@}

\def\fnum@table{\tablename~{\bf\thetable}}
\def\fnum@figure{\figurename~{\bf\thefigure}}
\def\tablename{\footnotesize{\bf Table}}
\def\figurename{\footnotesize{\bf Figure}}

\makeatother

\begin{document}

\title{\textbf{\huge The NE}\textbf{\Huge X}\textbf{\huge US Model}\huge }

\author{\textbf{K. Werner}\protect\( ^{1,\, a}\protect \)\textbf{, H.J. Drescher\protect\( ^{1,4}\protect \),
S. Ostapchenko\protect\( ^{1,2,3}\protect \), T. Pierog\protect\( ^{1}\protect \) }}

\date{\protect\( \qquad \protect \)}

\maketitle
{\par\centering \textit{\( ^{1} \)} \textit{\small SUBATECH, Université de Nantes
-- IN2P3/CNRS -- EMN,  Nantes, France }\\
\textit{\small \( ^{2} \) Moscow State University, Institute of Nuclear Physics,
Moscow, Russia}\\
 \textit{\small \( ^{3} \) Institute f. Kernphysik, Forschungszentrum Karlsruhe,
Karlsruhe, Germany}\\
\textit{\small \( ^{4} \) Physics Department, New York University, New York,
USA} \\
 \par}

\vspace{1cm}
{\par\centering \( ^{a} \) Invited speaker \par}

{\par\centering at the International Workshop on the Physics of the Quark Gluon
Plasma,\par}

{\par\centering Palaiseau, France, September 4-7, 2001\par}

{\par\centering  \par}

\vspace{1cm}
{\par\centering Abstract\par}

\begin{quote}
{\small The interpretation of experimental results at RHIC and in the future
also at LHC requires very reliable and realistic models. Considerable effort
has been devoted to the development of such models during the past decade, many
of them being heavily used in order to analyze data. }{\small \par}

{\small There are, however, serious inconsistencies in the above-mentioned approaches.
In this paper, we will introduce a fully self-consistent formulation of the
multiple-scattering scheme in the framework of a Gribov-Regge type effective
theory. }{\small \par}
\end{quote}
With the start of the RHIC program to investigate nucleus-nucleus collisions
at very high energies, there is an increasing need of computational tools in
order to provide a clear interpretation of the data. The situation is not satisfactory
in the sense that there exists a nice theory (QCD) but we are not able to treat
nuclear collisions strictly within this framework, and on the other hand there
are simple models, which can be applied easily but which have no solid theoretical
basis. A good compromise is provided by effective theories, which are not derived
from first principles, but which are nevertheless self-consistent and calculable.
A candidate seems to be the Gribov-Regge approach, and -- being formally quite
similar -- the eikonalized parton model. Here, however, severe inconsistencies
occur, as discussed in \cite{dre00}. 

As a solution of the above-mentioned problems, we present a new approach which
we call ``\textbf{Parton-based Gribov-Regge Theory}'': we have a consistent
treatment for calculating (topological) cross sections and particle production
considering energy conservation in both cases; in addition, we introduce hard
processes in a natural way. 

The basic guideline of our approach is theoretical consistency. We cannot derive
everything from first principles, but we use rigorously the language of field
theory to make sure not to violate basic laws of physics, which is easily done
in more phenomenological treatments (see discussion above).

\begin{figure}[htb]
{\par\centering \resizebox*{!}{0.125\textheight}{\includegraphics{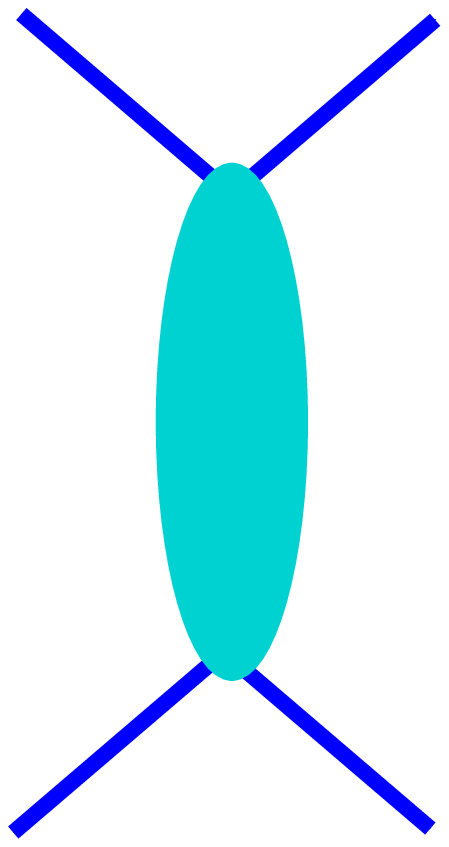}}  \( \qquad  \)\resizebox*{!}{0.125\textheight}{\includegraphics{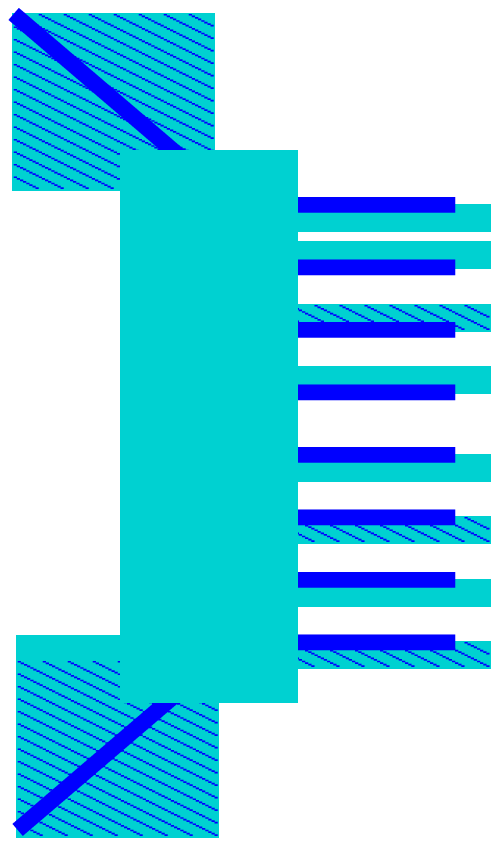}}  \par}

\caption{An elastic scattering amplitude \protect\( T_{2\rightarrow 2}\protect \) (left)
and an inelastic amplitude \protect\( T_{2\rightarrow X}\protect \) (right).\label{t}}
\end{figure}

\begin{figure}[htb]
{\par\centering \resizebox*{!}{0.12\textheight}{\includegraphics{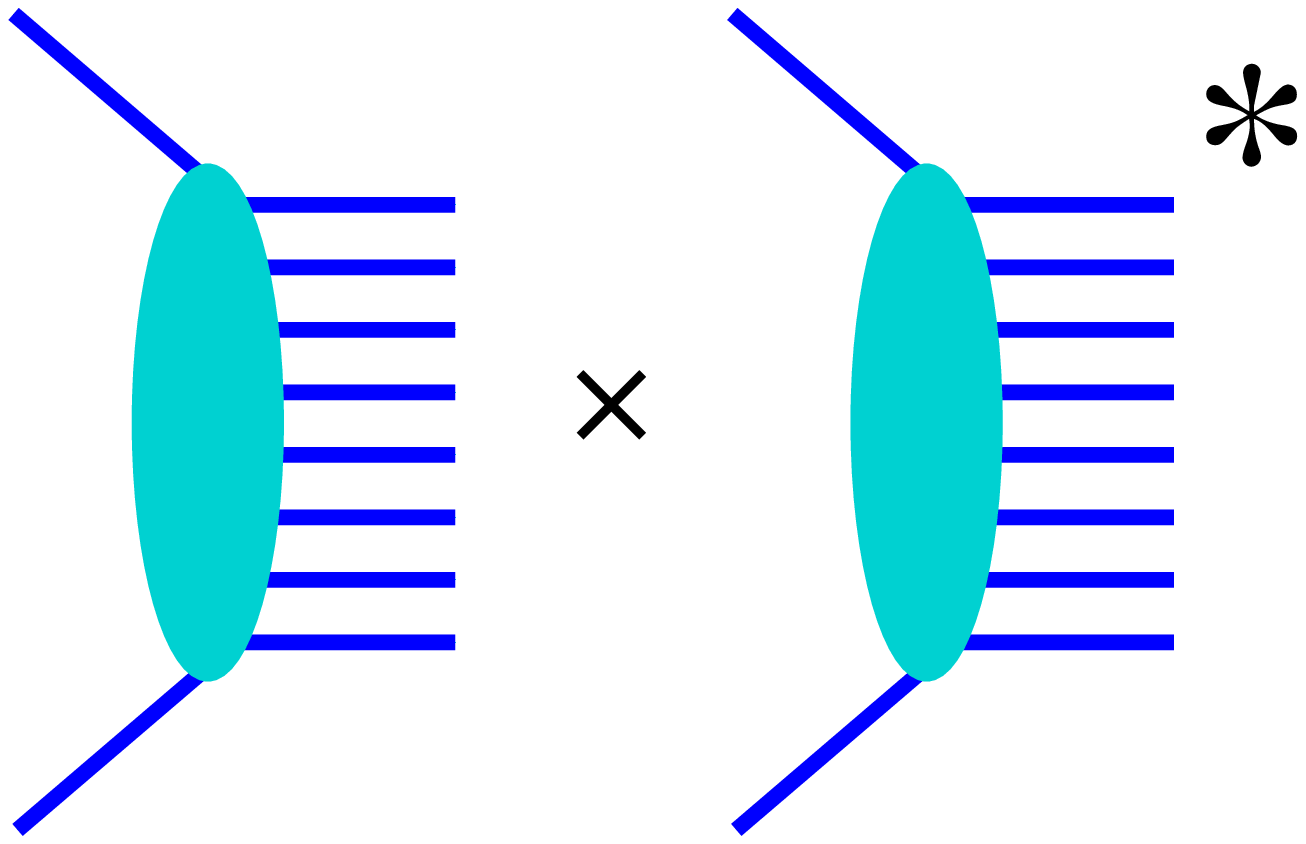}}  \resizebox*{!}{0.12\textheight}{\includegraphics{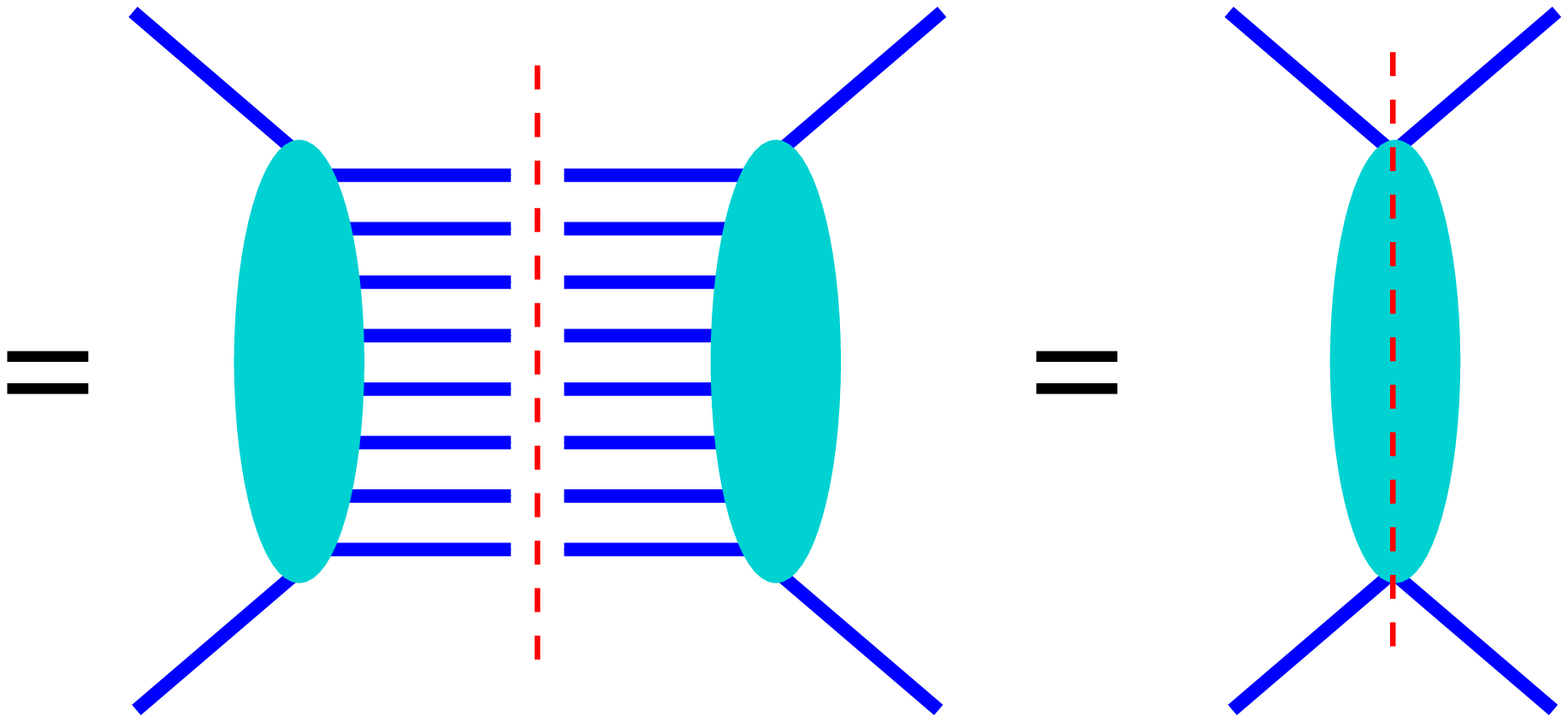}} \par}

\caption{The expression \protect\( \sum _{X}(T_{2\rightarrow X}).\protect \)\protect\( (T_{2\rightarrow X})^{*}\protect \)which
may be represented as a ``cut diagram''.\label{cut}}
\end{figure}

Let us first introduce some conventions. We denote elastic two body scattering
amplitudes as \( T_{2\rightarrow 2} \) and inelastic amplitudes corresponding
to the production of some final state \( X \) as \( T_{2\rightarrow X} \)
(see fig. \ref{t}). 
As a direct consequence of unitarity on may write the optical theorem  \( 2\mathrm{Im}T_{2\rightarrow 2}=\sum _{x} \)\( (T_{2\rightarrow X}) \)\( (T_{2\rightarrow X})^{*} \).
The right hand side of this equation may be literally presented as a ``cut
diagram'', where the diagram on one side of the cut is \( (T_{2\rightarrow X}) \)
and on the other side \( (T_{2\rightarrow X})^{*} \), as shown in fig. \ref{cut}. 
So the term ``cut diagram'' means nothing but the square of an inelastic amplitude,
summed over all final states, which is equal to twice the imaginary part of
the elastic amplitude.

\begin{figure}[htb]
{\par\centering \resizebox*{!}{0.15\textheight}{\includegraphics{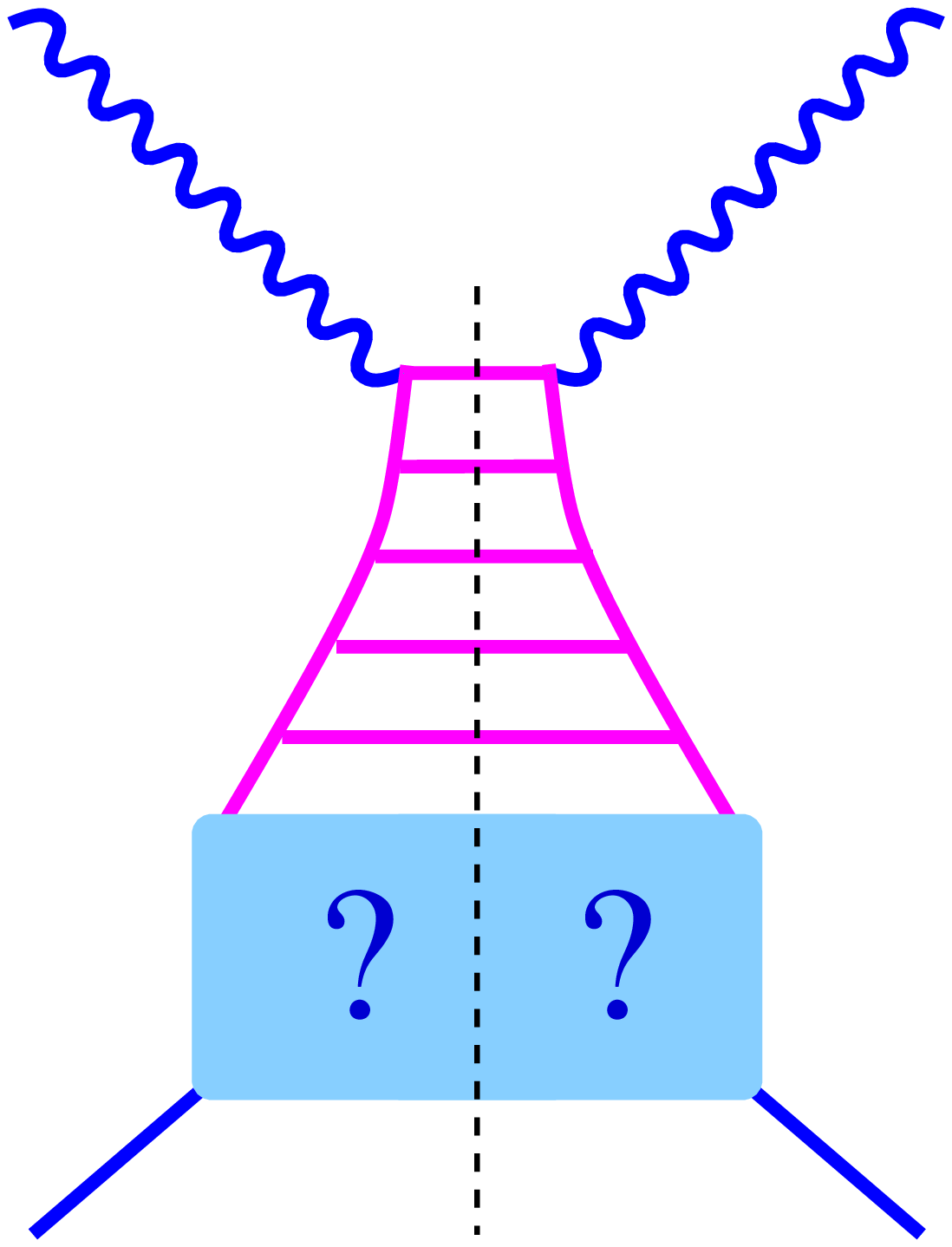}} 
\resizebox*{!}{0.15\textheight}{\includegraphics{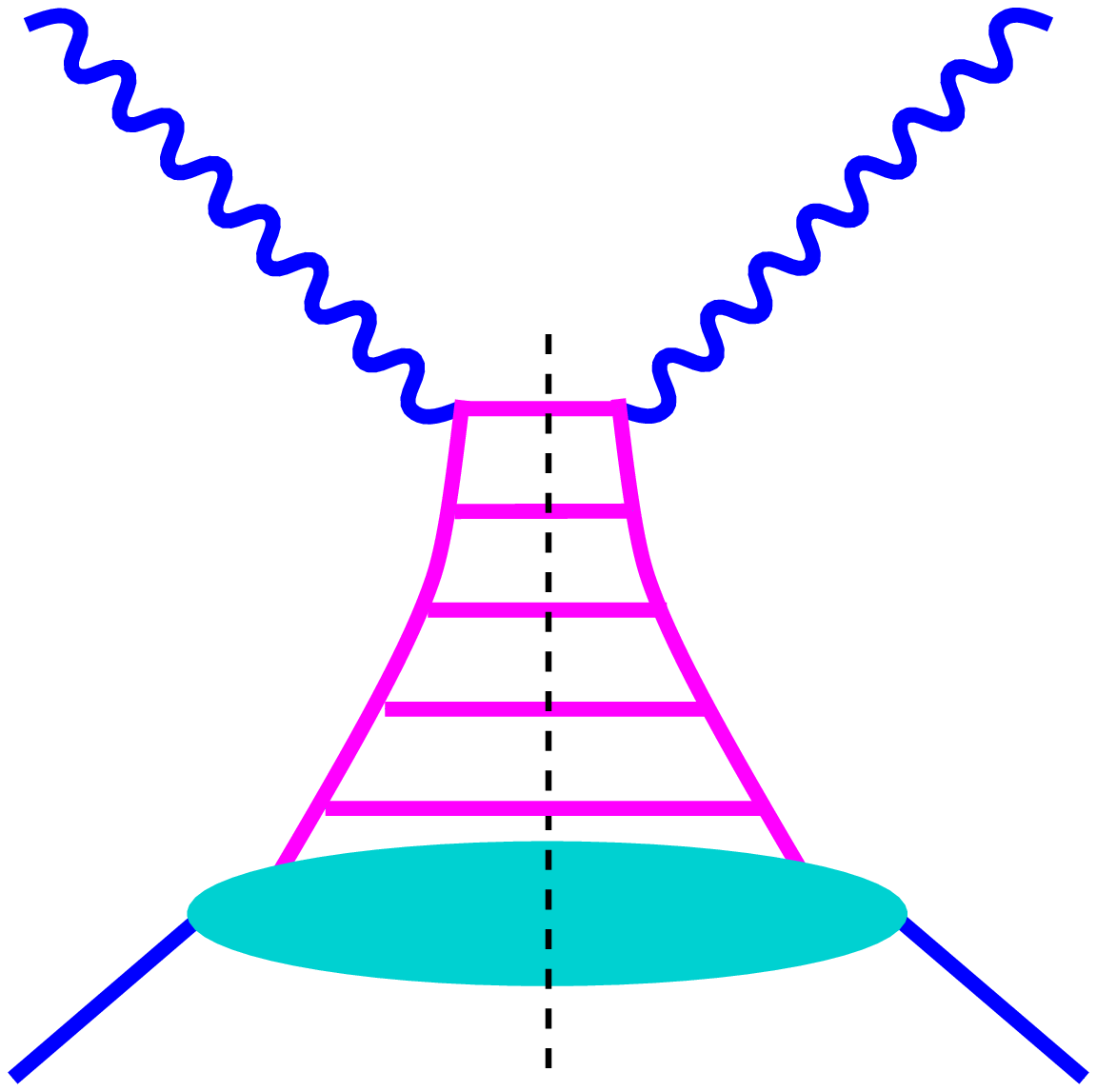}}  \resizebox*{!}{0.15\textheight}{\includegraphics{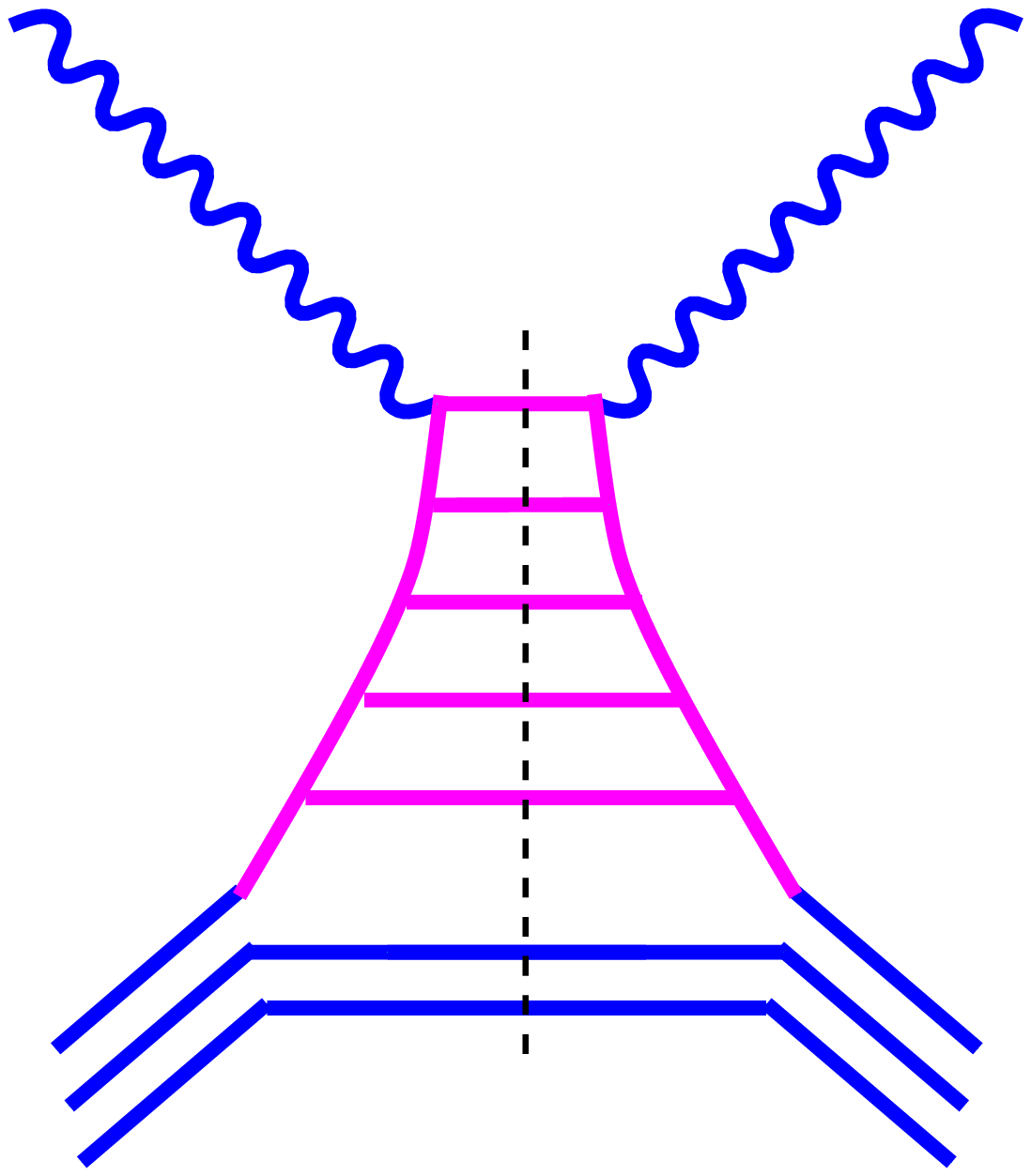}} \par}

\caption{Deep inelastic scattering: the general diagram (left) the ``sea'' contribution
with a soft Pomeron at the lower end (middle) and the ``valence'' contribution
(right).\label{dis}}
\end{figure}

Before coming to nuclear collisions, we need to discuss somewhat the structure
of the nucleon, which may be studied in deep inelastic scattering -- so essentially
the scattering of a virtual photon off a nucleon, see fig. \ref{dis}.
Let us assume a high virtuality photon. It is known that this photon couples
to a high virtuality quark, which is emitted from a parton with a smaller virtuality,
the latter on again being emitted from a parton with lower virtuality, and so
on. We have a sequence of partons with lower and lower virtualities, the closer
one gets to the proton. At some stage, some ``soft scale'' scale \( Q_{0}^{2} \)
must be reached, beyond which perturbative calculations are no longer valid.
So we have some ``unknown object'' -- indicated by ``?'' in fig. \ref{dis}
-- between the first parton and the nucleon. In order to proceed, one may estimate
the squared mass of this ``unknown object'', and one obtains doing simple
kinematics the value \( Q_{0}^{2}/x \), where \( x \) is the momentum fraction
of the first parton relative to the nucleon. Therefore, sea quarks or gluons,
having typically small \( x \), lead to large mass objects -- which we identify
with soft Pomerons, whereas valence quarks lead to small mass objects, which
we simply ignore. So we have two contributions, as shown in fig. \ref{dis}:
a sea contribution, where the sea quark or gluon is emitted from a soft Pomeron,
and a valence contribution, where the valence quark is one of the three quarks
of the nucleon. The precise microscopic structure of the soft Pomeron not being
known, it is parameterized in the usual way a a Regge pole.

\begin{figure}[htb]
{\par\centering \resizebox*{!}{0.11\textheight}{\includegraphics{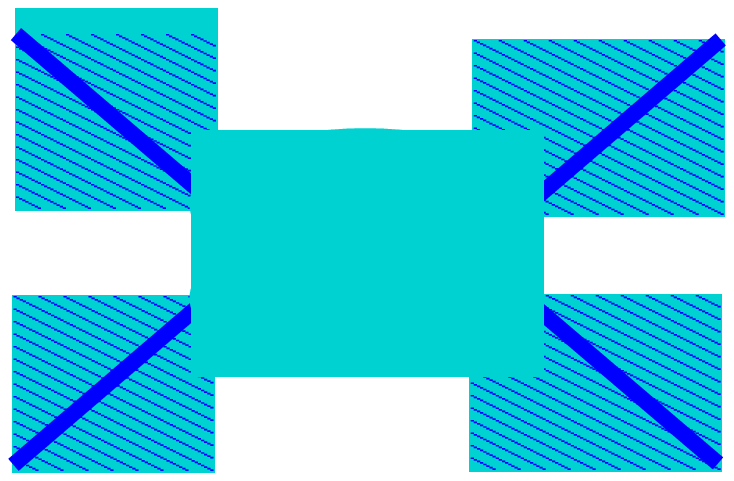}} \resizebox*{!}{0.15\textheight}{\includegraphics{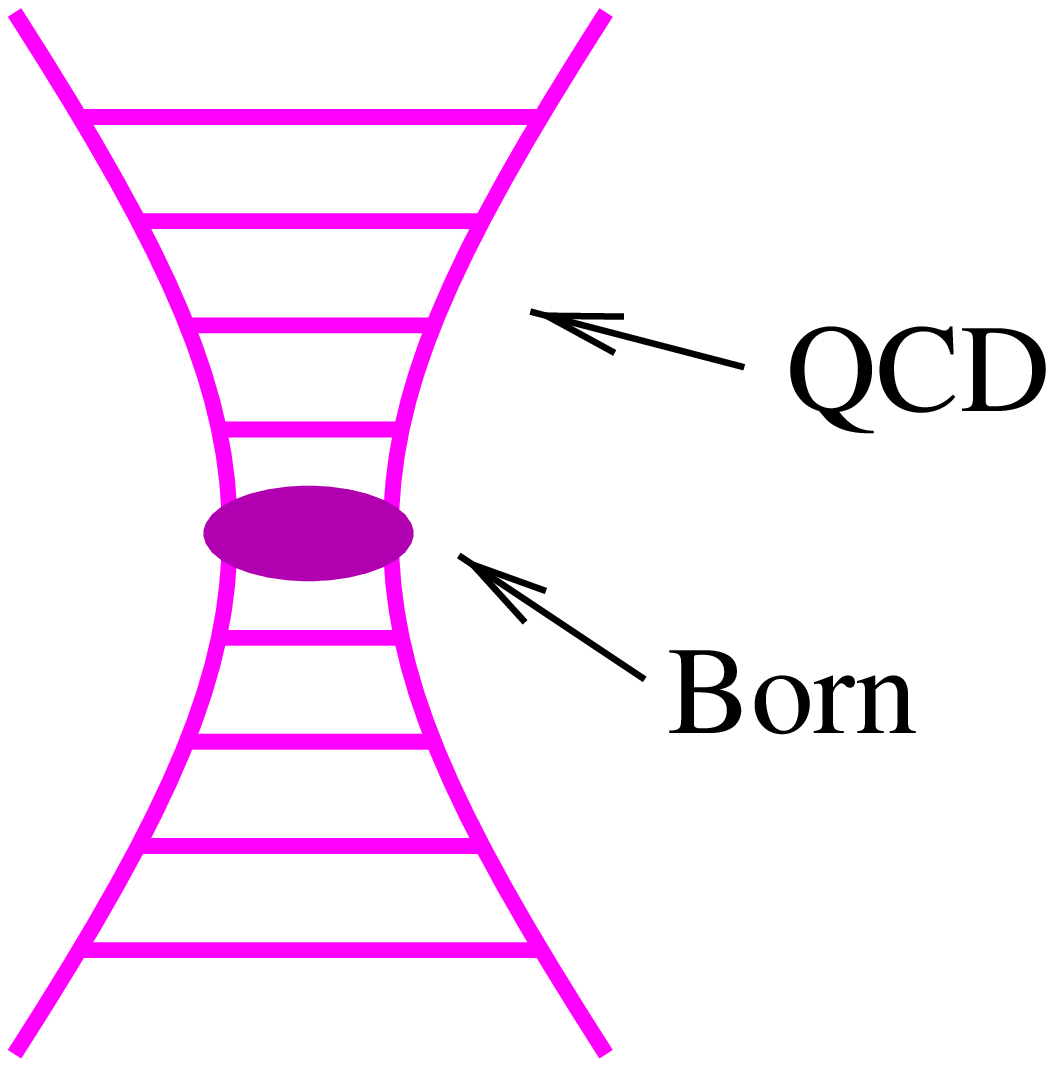}} \resizebox*{!}{0.15\textheight}{\includegraphics{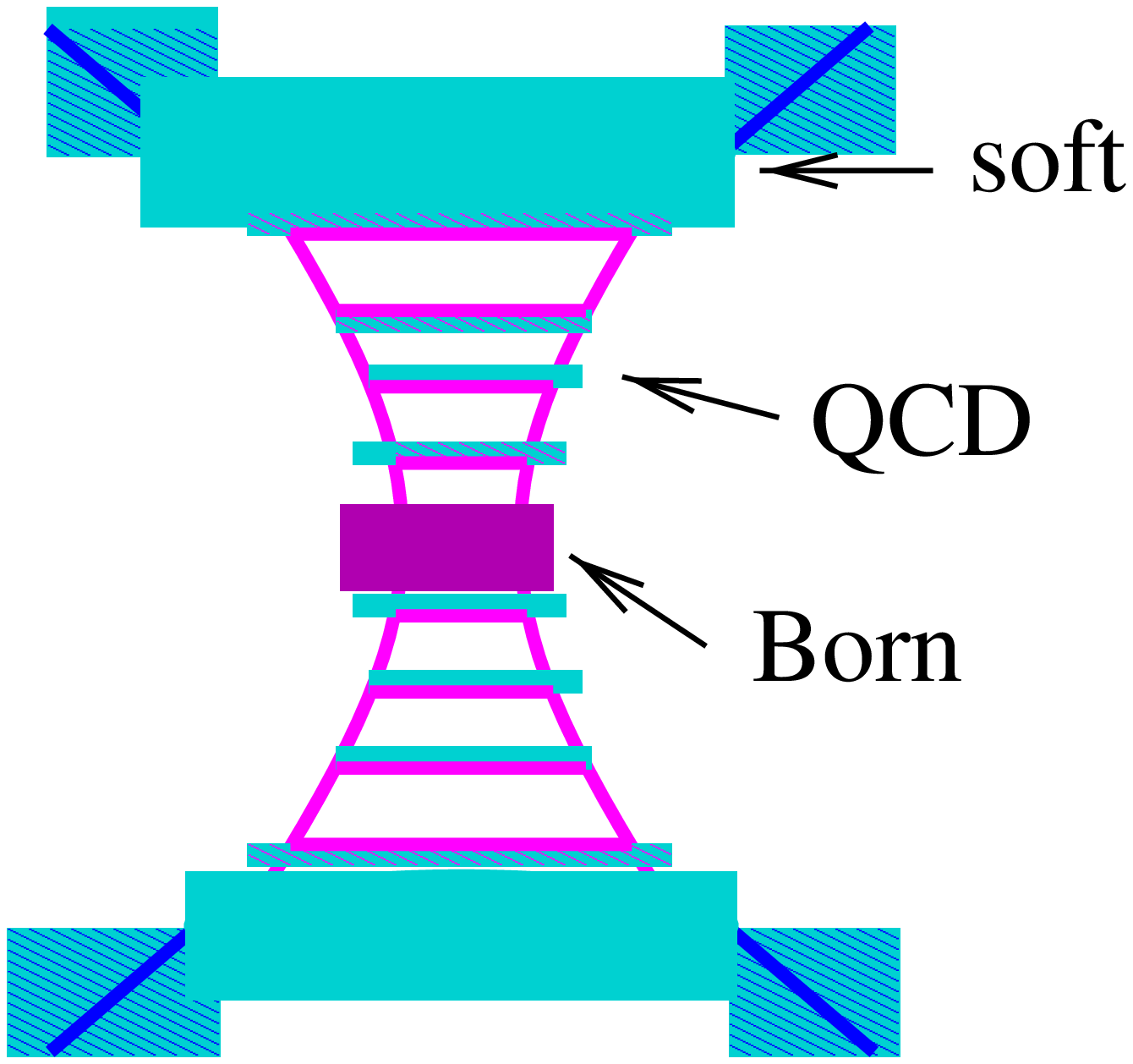}} \par}

\caption{The soft elastic scattering amplitude \protect\( T_{\mathrm{soft}}\protect \)
(left), the hard elastic scattering amplitude \protect\( T_{\mathrm{hard}}\protect \)
(middle) and one of the three contributions to the semi-hard elastic scattering
amplitude \protect\( T_{\mathrm{semi}}\protect \) (right). \label{nn}}
\end{figure}

Elementary nucleon-nucleon scattering can now be considered as a straightforward
generalization of photon-nucleon scattering: one has a hard parton-parton scattering
in the middle, and parton evolutions in both directions towards the nucleons.
We have a hard contribution \( T_{\mathrm{hard}} \) when the the first partons
on both sides are valence quarks, a semi-hard contribution \( T_{\mathrm{semi}} \)
when at least on one side there is a sea quark (being emitted from a soft Pomeron),
and finally we have a soft contribution, when there is no hard scattering at
all (see fig. \ref{nn}). 
We have a smooth transition from soft to hard physics: at low energies the soft
contribution dominates, at high energies the hard and semi-hard ones, at intermediate
energies (that is where experiments are performed presently) all contributions
are important.

Let us consider nucleus-nucleus (\( AB \)) scattering. In the Glauber-Gribov
approach \cite{gla59,gri69}, the nucleus-nucleus scattering amplitude is defined
by the sum of contributions of diagrams, corresponding to multiple elementary
scattering processes between parton constituents of projectile and target nucleons.
These elementary scatterings are exactly as discussed above, namely the sum
of soft, semi-hard, and hard contributions: \( T_{2\rightarrow 2}=T_{\mathrm{soft}}+T_{\mathrm{semi}}+T_{\mathrm{hard}} \).
A corresponding relation holds for the inelastic amplitude \( T_{2\rightarrow X} \).
We use the above definition of a cut elementary diagram, which is graphically
represented by a vertical dashed line, whereas the elastic amplitude is represented
by an unbroken line:

\vspace{0.3cm}
{\par\centering \resizebox*{!}{0.05\textheight}{\includegraphics{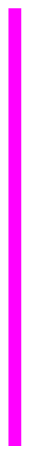}} \( \quad  \)\( =T_{2\rightarrow 2} \),
\( \quad  \)\( \quad  \)\resizebox*{!}{0.05\textheight}{\includegraphics{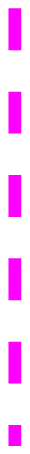}} \( \quad  \)\( =\sum _{X}(T_{2\rightarrow X}) \)\( (T_{2\rightarrow X})^{*} \).\par}
\vspace{0.3cm}

\begin{figure}[htb]
{\par\centering \resizebox*{!}{0.24\textheight}{\includegraphics{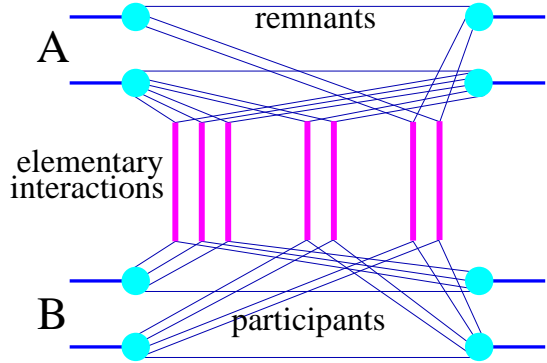}} \par}

\caption{Elastic nucleus-nucleus scattering amplitude, being composed of purely parallel
elementary interactions between partonic constituents of the nucleons (blobs).\label{aa}}
\end{figure}

\begin{figure}[htb]
{\par\centering \resizebox*{!}{0.24\textheight}{\includegraphics{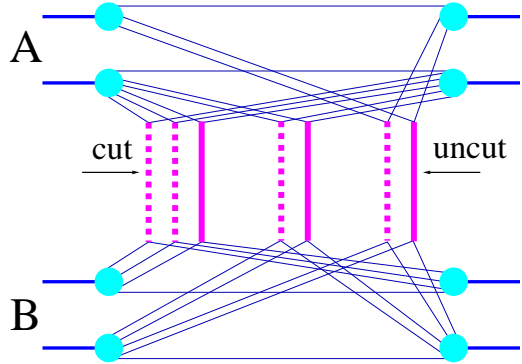}} \par}

\caption{Typical interference term contributing to the squared inelastic amplitude.\label{grtppaac}}
\end{figure}

\noindent This is very handy for treating the nuclear scattering model. We define
the model via the elastic scattering amplitude \( T_{AB\rightarrow AB} \) which
is assumed to consist of purely parallel elementary interactions between partonic
constituents, as discussed above. The amplitude is therefore a sum of terms
as the one shown in fig. \ref{aa}.
One has to be careful about energy conservation: all the partonic constituents
(lines) leaving a nucleon (blob) have to share the momentum of the nucleon.
So in the explicit formula one has an integration over momentum fractions of
the partons, taking care of momentum conservation. Having defined elastic scattering,
inelastic scattering and particle production is practically given, if one employs
a quantum mechanically self-consistent picture. Let us now consider inelastic
scattering: one has of course the same parallel structure, just some of the
elementary interactions may be inelastic, some elastic. The inelastic amplitude
being a sum over many terms -- \( T_{AB\rightarrow X}=\sum _{i}T^{(i)}_{AB\rightarrow X} \)
-- has to be squared and summed over final states in order to get the inelastic
cross section, which provides interference terms \( \sum _{X}(T^{(i)}_{AB\rightarrow X})(T^{(j)}_{AB\rightarrow X})^{*} \),
which can be conveniently expressed in terms of the cut and uncut elementary
diagrams, as shown in fig. \ref{grtppaac}. So we are doing nothing more than
following basic rules of quantum mechanics.
Of course a diagram with 3 inelastic elementary interactions does not interfere
with the one with 300, because the final states are different. So it makes sense
to define classes \( K \) of interference terms (cut diagrams) contributing
to the same final state, as all diagrams with a certain number of inelastic
interactions and with fixed momentum fractions of the corresponding partonic
constituents. One then sums over all terms within each class \( K \), and obtains
for the inelastic cross section

\noindent 
\[
\sigma _{AB}(s)=\int d^{2}b\: \sum _{K}\Omega ^{(s,b)}(K)\]
 where we use the symbolic notation \( d^{2}b=\int d^{2}b_{0}\, \int d^{2A}b_{A}\, \rho (b_{A})\, \int d^{2B}b_{B}\, \rho (b_{B}) \)
which means integration over impact parameter \( b_{0} \) and in addition averaging
over nuclear coordinates for projectile and target. The variable \( K \) is
characterized by \( AB \) numbers \( m_{k} \) representing the number of cut
elementary diagrams for each possible pair of nucleons and all the momentum
fractions \( x^{+} \) and \( x^{-} \) of all these elementary interactions
(so \( K \) is a partly discrete and partly continuous variable, and \( \sum  \)
is meant to represent \( \sum \int  \)). This is the really new and very important
feature of our approach: we keep explicitly the dependence on the longitudinal
momenta, assuring energy conservation at any level of our calculation. 

The calculation of \( \Omega  \) actually very difficult and technical, but
it can be done and we refer the interested reader to the literature \cite{dre00}. 

The quantity \( \Omega ^{(s,b)}(K) \) can now be interpreted as the probability
to produce a configuration \( K \) at given \( s \) and \( b \). So we have
a solid basis for applying Monte Carlo techniques: one generates configurations
\( K \) according to the probability distribution \( \Omega  \) and one may
then calculate mean values of observables by averaging Monte Carlo samples.
The problem is that \( \Omega  \) represents a very high dimensional probability
distribution, and it is not obvious how to deal with it. We decided to develop
powerful Markov chain techniques \cite{hla98} in order to avoid to introduce
additional approximations. 

To summarize: We provide a new formulation of the multiple scattering mechanism
in nucleus-nucleus scattering, where the basic guideline is theoretical consistency.
We avoid in this way many of the problems encountered in present day models.
We also introduce the necessary numerical techniques to apply the formalism
in order to perform practical calculations.

This work has been funded in part by the IN2P3/CNRS (PICS 580) and the Russian
Foundation of Basic Researches (RFBR-98-02-22024). 

\bibliographystyle{pr2}
\bibliography{a}

\end{document}